# The dielectric functions and optical band gaps of thin films of amorphous and fcc $Mg_{\sim2}NiH_{\sim4}$

J. H. Selj[*], T. Mongstad[*], B. Hauback, and S. Karazhanov

Institute for Energy Technology, P.O. Box 40, NO-2027 Kjeller, Norway

Magnesium nickel hydride films have earlier been suggested for several optoelectronic applications, but the optical properties and band gap have not been firmly established. In this work, the dielectric functions and the optical band gaps of thin films of $Mg_2NiH_4$ have been determined experimentally from optical modeling using spectroscopic ellipsometry and spectrophotometry in the photon energy range between 0.7 and 4.2 eV. Samples were prepared by reactive sputtering, resulting in a single-layer geometry that could easily be studied by ellipsometry. Crystalline samples were prepared by annealing the amorphous films *ex-situ*. The resulting films remained in the high temperature cubic $Mg_2NiH_4$ structure even after cooling to room temperature. Tauc analysis of the dielectric function shows that $Mg_2NiH_4$ exhibits a band gap of 1.6 eV for amorphous structure films and 2.1 eV for its fcc crystalline structure.

Keywords: Metal hydride, magnesium nickel hydride, thin film, ellipsometry, optical properties, structural properties, crystallinity, band gap estimation

---

[*] Corresponding authors. E-mail adresses: josefine.selj@ife.no (J. Selj), trygve.mongstad@ife.no (T. Mongstad).

# 1 Introduction

Many metals become semiconductors or insulators upon hydrogen absorption, forming metal hydrides. This transition in metal hydrides has been known for decades, but it first caught significant attention when it was discovered in 1996 that this effect could be utilized in thin-film metal hydride smart windows [1]. While the first metal hydride switchable windows were based on yttrium and lanthanum, the attention was soon drawn to Mg- and Ni-based hydrides because of the high abundance of these materials [2]. Recently, semiconducting $Mg_2NiH_4$ has also been suggested as a potential material for electronics and photovoltaics [3] and digital memory [4].

When alloys of Mg and Ni are hydrogenated, a semiconducting metal hydride is formed. The stoichiometric compounds are $MgH_2$ and $Mg_2NiH_4$, but any composition of $Mg_yNi$ with $y >\sim 0.5$ will partly or fully react with H and form a metal hydride [5]. The optical and electrical properties are highly dependent on the composition $y$. Even though the optical properties of $Mg_yNiH_x$ is central to many of the suggested technological applications, only a few reports of the dielectric functions of $Mg_{\sim 2}NiH_{\sim 4}$ are found in the literature [6,7].

There is a multitude of different reports of the band gap of $Mg_2NiH_4$, ranging from $E_g \approx 1.1$ eV to $E_g \approx 2.2$ eV [5,7–12]. While some of the variation in the band gap is related to the uncertainties in the composition and the synthesis method of the films, the main reasons for the variations are probably the fitting of the experimental optical data and the use of different methods to assign the band gap.

In this work we have produced thin films of semiconducting amorphous $a$-$Mg_{\sim 2}NiH_{\sim 4}$ and crystalline $c$-$Mg_{\sim 2}NiH_{\sim 4}$; the two phases that are possible to form as thin films. The films have been characterized by ellipsometry and spectrophotometry to establish the dielectric functions and to evaluate their optical band gaps. The well-established band gap estimation methods of Tauc [13] have been utilized to determine the band gap.

# 2 Experimental details

## 2.1 Thin film synthesis

Thin film deposition was performed by reactive co-sputtering of metallic Mg and Ni targets with a Leybold Optics A550V7 inline sputtering system. The purity of the targets was 99.5% and 99.8%, respectively. The target area was 125 x 600 mm$^2$, and the targets were set at an angle of 30° with respect to the substrate against each other in order to enhance the co-sputtering. The mean distance from the targets to the substrate was 116 mm, with center-to-center distance between the targets of 210 mm. The Mg target was operated with 800 W RF (13.56 MHz) power and the Ni target with 200 W DC power. The co-sputtering yielded films with a spatial gradient in the composition, and thus making it easy to scan over the sample to obtain optical data for different Mg-Ni compositions. The purity of the gases used during sputter deposition was 5N for Ar and 6N for $H_2$. The base pressure of the chamber was $1.6 \times 10^{-4}$ Pa, and the depositions were performed under 0.4 Pa pressure. A 3:7 mixing ratio of $H_2$ to Ar was used, with a total gas flow of 200 sccm. The samples were deposited on 76 × 26 mm$^2$ glass substrates (Menzel-Gläser microscope slides), cleaned in an ultrasound bath of de-ionized water prior to the deposition. The depositions were carried out at room temperature, giving amorphous metal hydride films. To compare the optical properties of amorphous and crystalline samples, crystalline samples were formed by annealing *ex-situ* at a temperature of 523 K for 30 min after deposition.

## 2.2 Sample characterization

The films have been characterized by ellipsometry and spectrophotometry to establish the dielectric functions and to evaluate their optical band gaps. Ellipsometry measurements were conducted using a Woollam variable angle spec-troscopic ellipsometer (VASE) in the wavelength range 300–1700 nm. To obtain adequate sensitivity over the whole spectral range, two angles of incidence (70° and 75°) were used. The diameter of the beam spot was 0.4 cm.

The absence of the catalytic Pd cap layer, which is commonly used in synthesis of thin-film metal hydrides, simplified our optical modeling and the ellipsometric measurements as the ellipsometry could be performed from the front side of the sample. A diffusely scattering polymer adhesive tape was attached to the back side of the glass substrate to avoid collecting light reflected from the rear side. The VASE ellipsometer was also used for collecting reflectance and transmittance data. The combination of ellipsometric data and spectrophotometry, i.e. calculation of the dielectric functions from measurements of optical reflection and transmission, gives a more accurate and robust optical analysis than spectrophotometry or ellipsometry alone. All ellipsometric data were analyzed with the J.A. Woollam WVASE32 software package.

Structural characterization was performed by x-ray diffraction (XRD) in a Bruker D8 Discover diffractometer with Cu-Kα radiation. Grazing incidence XRD (GI-XRD) was used with an incident angle of 0.5°. The Mg-Ni composition was measured by Rutherford backscattering spectrometry (RBS) and energy-dispersive x-ray spectroscopy (EDS) on the as-deposited films. The H content was measured using the $N^{15}$ nuclear reaction analysis (NRA) method [14]. The NRA and RBS measurements were carried out at Tandem Accelerator Laboratory at Uppsala University, Sweden, and analyzed using the SIMNRA software [15].

# 3 Theory and background

## 3.1 Ellipsometry

The dielectric properties of the glass substrate were determined prior to the deposition of the metal hydride films using ellipsometry and transmission data. The dielectric function of $Mg_{\sim 2}NiH_{\sim 4}$ films were obtained by fitting the measured ellipsometric data using a linear combination of dispersion functions (oscillators) describing the dielectric function over the measured spectral range. The oscillator parameterization has the advantage that the fitted dielectric function is Kramers-Kronig consistent. In this work we are using Tauc-Lorentz (TL) and Gauss oscillators. The fit parameters of a Gaussian oscillator are the amplitude ($A$), the center energy ($E_c$), and the broadening ($B$). The TL oscillator developed by Jellison and Modine [16] is asymmetrical and was originally developed to fit amorphous materials. However, with its asymmetric shape it can also be used for curve fitting of band gaps and higher absorptions of crystalline materials since the absorption is zero below the gap and greater than zero above the gap. The TL oscillator is characterized by its amplitude, center energy, broadening, and an oscillator band gap ($E_{g,TL}$). If the dielectric function is described by a TL oscillator alone, this TL band gap would provide an estimate of the optical band gap. However, the TL oscillator is used together with a Gaussian and information about the optical band gap cannot be directly deduced from the oscillator parameters. An additional fitting parameter, $\varepsilon_\infty$, takes into account all excitations at higher energies than those included in the ellipsometry spectra. The oscillator parameters obtained for different compositions of $a$- and $c$-$Mg_yNiH_x$ are shown in Table 1. The fit to the ellipsometric variables $\varphi$ and $\Delta$ for a representative sample is shown in Fig. 1.

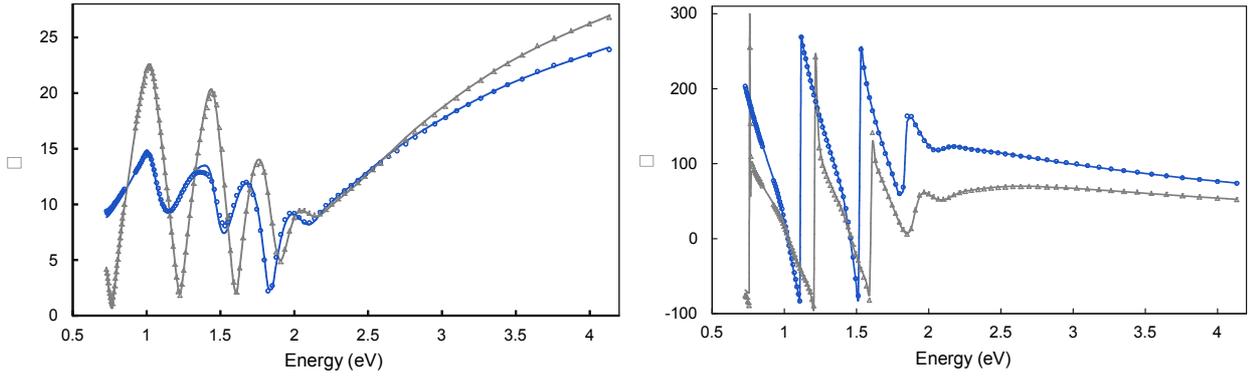

**Fig. 1.** The fit of the model to the ellipsometric variables $\varphi$ and $\Delta$ for a 509 nm thick $a$-Mg$_{2.1}$NiH$_x$ film on glass.

**Table 1.** Oscillator parameters of the best fit to the ellipsometry, transmittance and reflectance data for different compositions of $a$- and $c$-Mg$_y$NiH$_x$

| Sample | Oscillator | $A$ | $E_c$ | $B$ | $E_{g,TL}$ |
|---|---|---|---|---|---|
| $a$-Mg$_{2.3}$NiH$_x$ | 1 Gauss | 6.51 | 3.03 | 1.16 | |
| | 2 TL | 158.74 | 3.63 | 2.25 | 2.78 |
| $a$-Mg$_{2.1}$NiH$_x$ | 1 Gauss | 6.63 | 2.97 | 1.23 | |
| | 2 TL | 180.17 | 3.54 | 2.31 | 2.83 |
| $a$-Mg$_{1.9}$NiH$_x$ | 1 Gauss | 7.35 | 3.06 | 1.50 | |
| | 2 TL | 207.0 | 4.23 | 3.54 | 2.82 |
| $c$-Mg$_{2.4}$NiH$_x$ | 1 Gauss | 5.32 | 3.02 | 0.71 | |
| | 2 TL | 137.2 | 3.23 | 0.92 | 2.73 |
| $c$-Mg$_{2.2}$NiH$_x$ | 1 Gauss | 2.68 | 2.77 | 0.50 | |
| | 2 TL | 223.4 | 3.13 | 0.75 | 2.54 |
| $c$-Mg$_{2.0}$NiH$_x$ | 1 Gauss | 2.03 | 2.65 | 0.40 | |
| | 2 TL | 250.4 | 3.09 | 1.00 | 2.49 |

### 3.2 The band gap of Mg$_y$NiH$_x$

Several values of the band gap of Mg$_2$NiH$_4$ have been reported in the literature. The different values are due to several factors. The first is the existence of three different crystalline phases of Mg$_2$NiH$_4$, with different electronic properties. The two low-temperature phases (LT1, LT2) are modifications of a monoclinic structure [17] and the high-temperature phase (HT) has a cubic fcc crystal structure [18]. In addition, hydrogenating thin metallic films at room temperature gives an amorphous state, $a$-Mg$_2$NiH$_4$. Only $a$-Mg$_2$NiH$_4$ and HT cubic $c$-Mg$_2$NiH$_4$ structures are reported to form in thin films. Films that have been hydrogenated above 510 K remain in the HT-structure even when cooled to room temperature, contrary to bulk Mg$_2$NiH$_4$ [4]. The second most important reason for the confusion about the band gap is the choice of method to derive the optical constants and perform the band gap estimation. A third reason is the effect of the stoichiometry of Mg and Ni, in which the band gap can be changed with more than 0.5 eV [5].

The first published report claimed a band gap of 1.68 eV for both the monoclinic LT phase and the fcc HT phase [8]. These estimations were based on temperature-dependent measurements of electrical resistivity on pressed powders. Af-ter the discovery of the switchable window based on Mg and Ni [2], several experimental and computational studies on the optical properties of $Mg_2NiH_4$ were published. Using density functional theory (DFT), Myers et al. [9] found a band gap of 1.4 eV for LT-$Mg_2NiH_4$ and 1.9 eV for HT-$Mg_2NiH_4$. A lower gap of 1.17 eV for HT-$Mg_2NiH_4$ had earlier been suggested in a similar DFT study [19]. Experimental studies on $a$-$Mg_2NiH_4$ films have showed values of 1.9 eV [7], 1.8 eV [5], 1.75 eV [12], 1.6 eV [10] and 1.3 eV [11], while a band gap of 2.2 eV was recently estimated for a $c$-$Mg_2NiH_4$ film.

The most used method for estimation of the band gap from optical measurements is the one proposed by Tauc [13]. The optical band gap can be related to the frequency dependence of the absorption coefficient:

$$\alpha(v) = (hv - E_g)^m / hv$$

For amorphous and indirect band gap materials $m = 2$ is used. DFT calculations done on the cubic structure of $Mg_2NiH_4$ gives reason to expect an electronic structure with an indirect band gap [19]. The band gap is estimated by plotting $(\alpha hv)^{1/m}$ as a function of the photon energy $hv$ and extrapolating the linear region to zero.

# 4 Results and discussion

## 4.1 Synthesis and structural properties of the films

The most common structural form of thin-film $Mg_2NiH_4$ is the amorphous, which is the typical result after hydrogenation of Pd-capped metallic $Mg_2Ni$ films at room temperature [5] or films deposited *in-situ* by reactive deposition [12,20], both processes at room temperature. In this work we have used reactive co-sputtering to form the hydride films. Fig. 2(a) shows the x-ray scattering from different compositions of $a$-$Mg_yNiH_x$. The spectra demonstrate the same features for all compositions, with halos centered at $2\theta = 23°$ and $2\theta = 40°$. Although the films are found to be amorphous for XRD with Cu Kα radiation, our earlier work has showed that crystallites of monoclinic $Mg_2NiH_4$ are observable with transmission electron microscopy (TEM) [20]. The crystal size found with TEM was 5-10 nm [20].

Thin films of crystalline $Mg_2NiH_4$ have been prepared earlier by hydrogenation of Mg-Ni films above the phase transition temperature [4]. The films remained in the cubic high-temperature structure also after returning to room temperature. In this work we synthesized crystalline films by heating the samples *ex-situ* after deposition. The annealing gave close to identical results after heating in air, $N_2$ and $H_2$ atmospheres. The samples studied by XRD and ellipsometry in the current work have been crystallized by annealing the samples at 523 K for 30 minutes in air. Fig. 2(b) shows the diffraction patterns for various compositions of $Mg_yNiH_x$ thin films. The films take the HT fcc structure of $Mg_2NiH_4$ of space group Fm-3m and lattice parameter $a = 6.50$ Å [18]. The peak intensities are close to those expected for random orientation of the crystallites. There is no displacement of the peaks due to the variation in composition. Bragg peaks from metallic $Mg_2Ni$ is also visible at about 20° and 44°. Samples with lower Mg content than $y \approx 1$ did not become crystalline, as demonstrated from the spectra for $Mg_{1.0}NiH_x$ in Fig. 2(b).

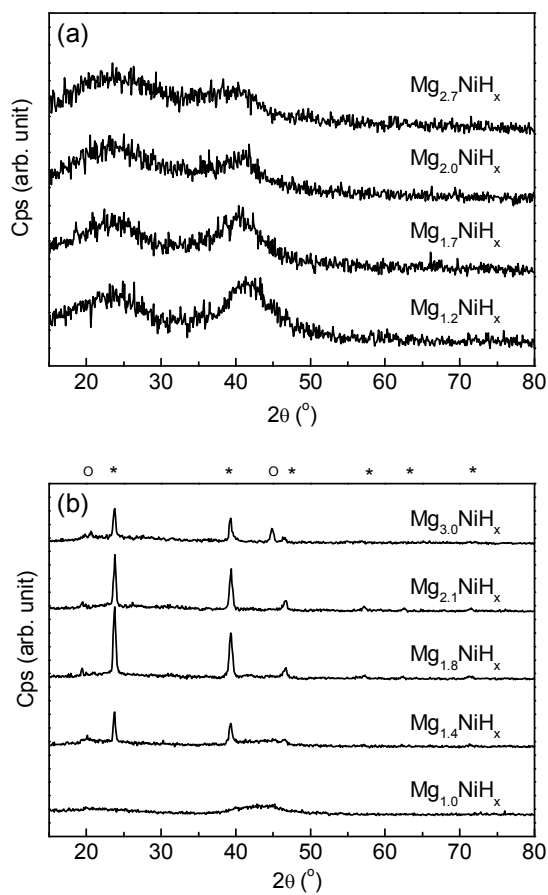

**Fig. 2.** XRD patterns from $Mg_yNiH_x$ films with different compositions. a) Films in the amorphous as-deposited state. b) Films in the crystallized fcc structure. Peaks originating from the cubic $Mg_2NiH_4$ structure are marked with (*), peaks from hexagonal $Mg_2Ni$ are marked with (°).

Fig. 3 shows elemental composition analysis for $Mg_2NiH_4$ films. The RBS measurement in Fig. 3(a) was done on an amorphous $Mg_2NiH_4$ film deposited on a carbon substrate. The data shows no sign of a gradient in the Mg or Ni composition through the thickness of the film. There are oxide layers both on the top of the carbon substrate and on the upper surface of the film, with thicknesses of ~5 nm and ~10 nm, respectively. From the RBS spectrum we estimate ~1% O contamination in the bulk of the film, but the uncertainty is high because the signal in this region (channel 130-190) is close to the background level. There could also be up to ~1% contamination of another impurity element with an atomic weight between Mg and Ni. Fig. 3(b) shows the H profile for the top 100 nm of one amorphous and one crystalline film, obtained by NRA. The graphic shows that the H is well-distributed through the film, and that there is no substantial change in the H content due to the crystallization annealing. The data has not been calibrated using a standard, and could thus not be used for estimation of the absolute H concentration. Therefore we choose to denote the H content of the films with $x$ in $Mg_yNiH_x$, although we expect them to have a H concentration of approximately two times the concentration of Mg ($x = 2y$ in $Mg_2NiH_4$ and $MgH_2$).

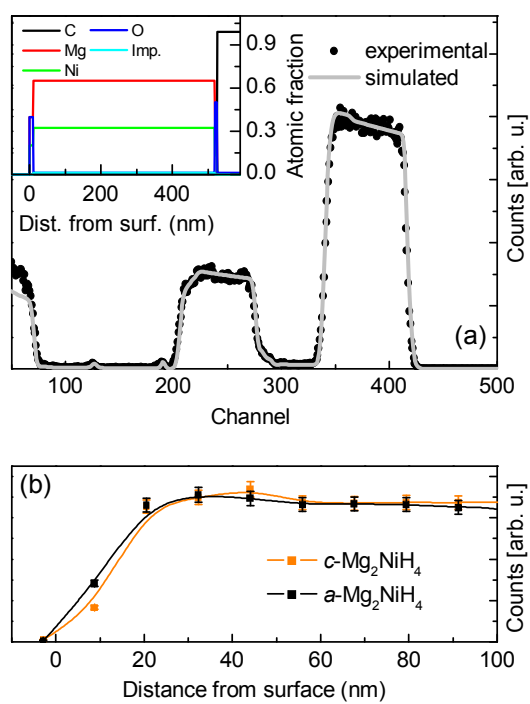

**Fig. 3.** Compositional analysis of thin films of $Mg_2NiH_4$. a) RBS data for a 500 nm thick $a$-$Mg_2NiH_4$ film and the fit to the data. The inset shows the corresponding profile of the atomic fractions excluding H. b) NRA H profile for crystalline and amorphous $Mg_2NiH_4$.

### 4.2 Optical properties and layering

Extracting the dielectric function of a thin film from ellipsometric data is significantly simplified and provide more robust results if some information about the film, such as thickness, layering or gradients, is known in advance. In the present work, an approximate thickness was found from profilometry and, as discussed in the previous section, structural and compositional information is obtained by RBS, TEM, and NRA.

From the modeling of the optical data we found that a model with a small variation of the dielectric function through the thickness of the film provided the best fit. Analysis by RBS and TEM showed that there is insignificant variation in the Mg and Ni concentration through the thickness of the samples [14]. NRA analysis similarly showed relatively constant concentrations of H. However, the temperature history of the hydride film can give rise to variation of the optical properties, as clearly illustrated by the structural and optical changes of $Mg_yNiH_x$ during annealing.

The optical gradient found for the amorphous samples corresponds to a variation of approximately 10% in the real and complex parts of the dielectric function. The gradient model was only marginally better than an ungraded model when only ellipsometric data was fitted. However, to obtain a good fit with the optical transmission data, incorporation of a gradient was necessary. For all samples the refractive index ($n$) and extinction coefficient ($k$) is gradually increasing from the substrate interface towards the surface of the film. Figure 4 shows the in-depth variation in the real part of the refractive index, $n$, for $a$-$Mg_{2.1}NiH_x$ as obtained by ellipsometric modeling. The gradients of the other amorphous samples are very similar. For the crystalline samples, the variation in the dielectric function is only a few percent. We believe that the gradient in the optical properties is due to temperature increase during the deposition process. The substrate temperature can increase from room temperature to 50-100 ºC during processing due to chemical reactions and the high kinetic energy of the deposited ions.

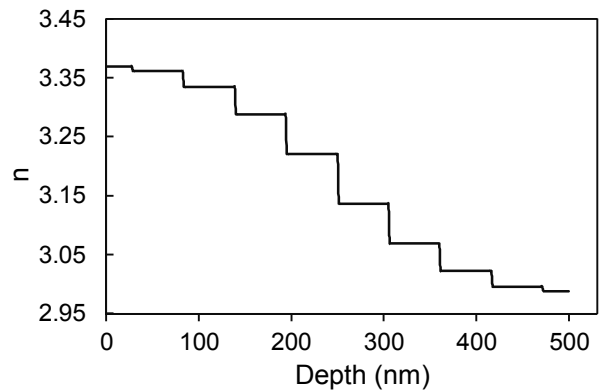

**Fig 4.** The gradient in the real part of the refractive index ($n$) for a 509 nm thick $a$-$Mg_{2.1}NiH_x$ sample as determined by ellipsometry. The percentual gradient in the imaginary part of the refractive index ($k$) is identical to the gradient in $n$.

Furthermore, a surface layer must be added in the model in order to provide a good fit to the ellipsometric data. For the amorphous sample, tabulated values for MgO [21] give a good fit to the data. For the crystalline films the surface layer is modeled by a Gaussian oscillator as this results in a significantly better fit than with the MgO surface layer. The modeled surface layers are between 7 and 11 nm thick, in accordance with the experimental results from the RBS analysis (Fig. 3). An intermediate oxide layer between the metal hydride film and the quartz substrate did not improve the fit to the data. This is not very surprising since a very thin intermediate oxide layer would be difficult to detect on a glass substrate due to the low refractive index contrast.

Fig. 5 displays the measured and fitted reflectance and transmission for ~500 nm thick $a$-$Mg_2NiH_x$ and $c$-$Mg_2NiH_x$ films. Good fits to both ellipsometric and spectrophotometric data were achieved for all samples

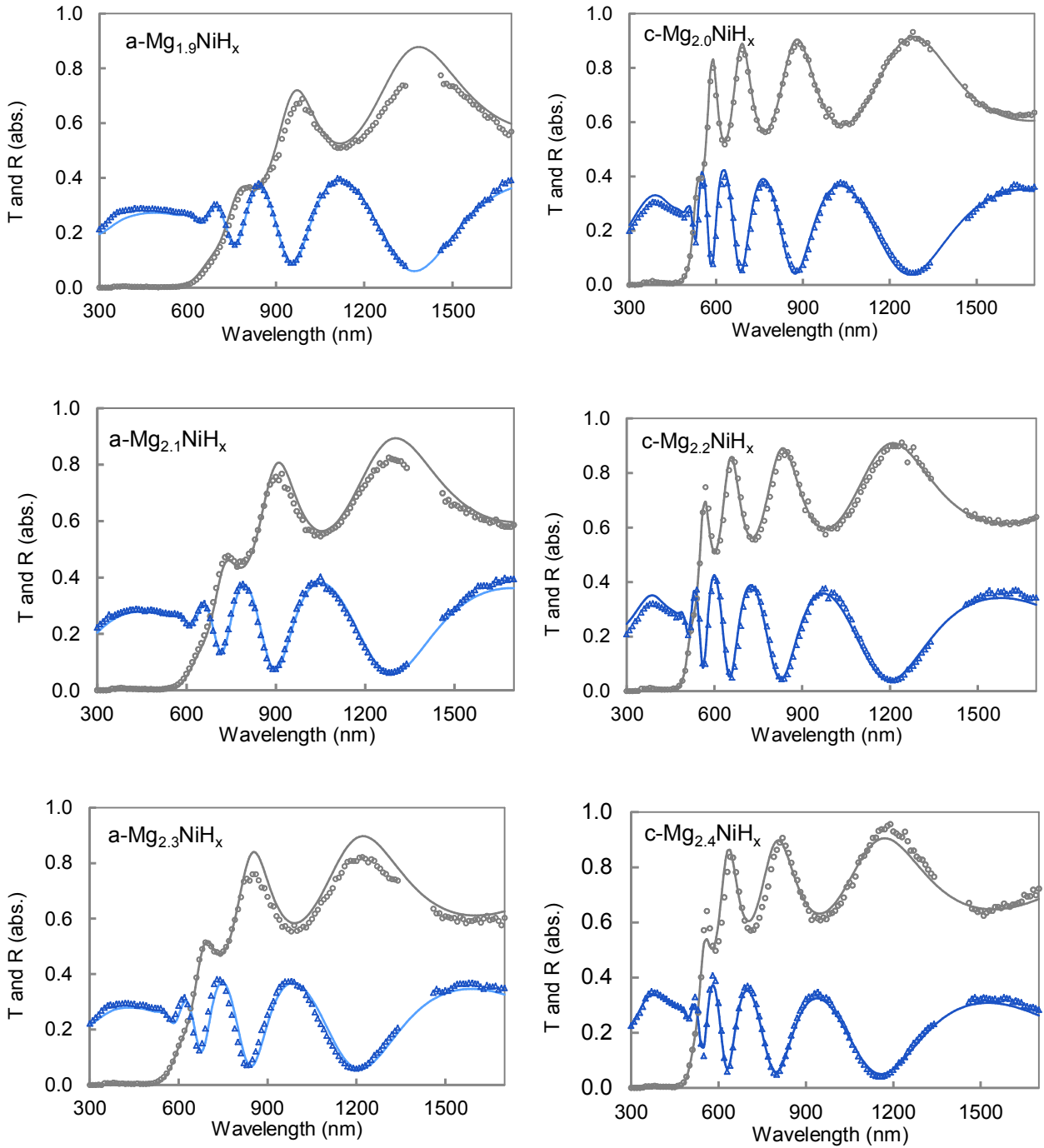

**Fig. 5.** The experimental reflectance (triangle) and transmittance (circles) and the calculated reflectance and transmittance resulting from the combined fit of ellipsometric-, reflectance-, and transmittance data for $a$-Mg$_y$NiH$_x$ and $c$-Mg$_y$NiH$_x$.

## 4.3 The dielectric functions determined by ellipsometry

### 4.3.1 Dielectric function of $a$-Mg$_y$NiH$_x$

Figure 6 illustrates the dielectric functions for $a$-Mg$_y$NiH$_x$ with $1.9 \leq y \leq 2.3$. The plotted dielectric functions are from the middle of each film. The dielectric functions for $a$-Mg$_y$NiH$_x$ with $y \approx 2$ have earlier been evaluated for Pd-capped hydrogenated samples by Ell et al. [6] and Lohstroh et al. [7]. Both fitted the dielectric function to optical reflection and transmission data, using Lorentz oscillators. The general features of the dielectric functions found in the current work resemble these earlier studies. Ell et al. could not achieve full hydrogenation for $y < 4$, and therefore modeled the data with three different optical layers modeled as effective media mixtures with up to 22% metallic particles of Mg$_2$Ni. Lohstroh et al. modeled the Mg$_2$NiH$_4$ film as a homogenous layer using four Lorentz and one Drude term in the model. The Drude term was included to account for metallic particle inclusions. In our work, ellipsometric modeling with inclusions of metallic particles did not improve the fit to the experimental data, and the high transparency below the band gap and high electrical resistivity suggests low presence of metallic domains in the films [20]. We have do not presented the data for samples with $y < 1.9$ and $y > 2.4$ because the optical absorption for these films was high below the band gap, and thus suggesting high presence of metallic particles.

### 4.3.2 Dielectric function of $c$-Mg$_y$NiH$_x$

Even though crystalline films of Mg$_2$NiH$_4$ have been synthesized earlier [4], the optical properties have not been thoroughly evaluated. Visually the crystallized sample appears yellow-transparent whereas the as-deposited amorphous sample appears darkish red. The dielectric functions of the annealed samples of $c$-Mg$_y$NiH$_x$ are shown in Fig. 7. This sample shows significantly narrower absorption peaks than the amorphous sample. The absorption approaches zero for low photon energies, indicating no metallic particles in the film.

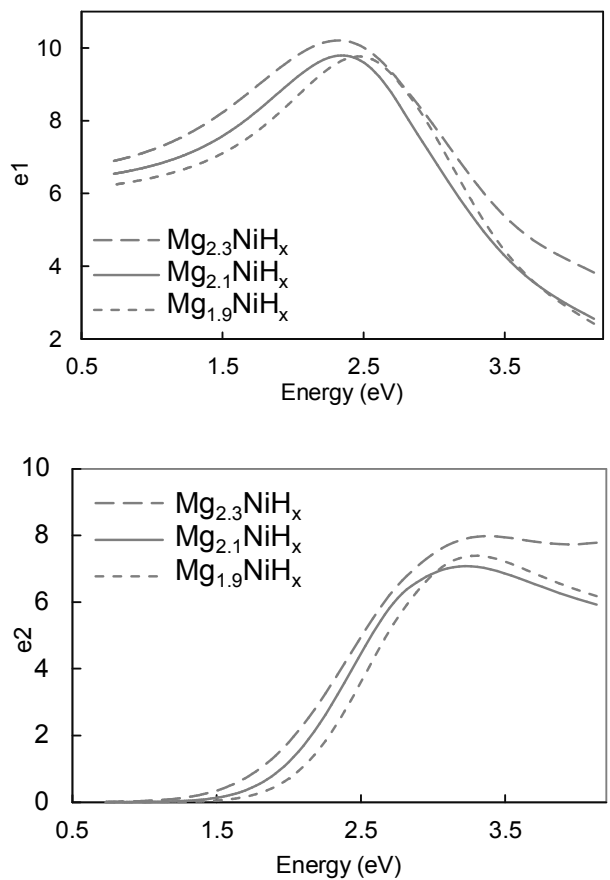

**Fig. 6.** The modelled dielectric functions determined from ellipsometry for $a$-Mg$_y$NiH$_x$.

Increasing the Mg contents leads to a decrease in the real part of the dielectric function for energies below 2.5 eV for both the amorphous and the crystalline samples.

### 4.4 The band gap of Mg$_2$NiH$_4$

We have used the Tauc method to estimate the band gap for $a$-Mg$_y$NiH$_x$ with $1.9 \leq y \leq 2.3$ and $c$-Mg$_y$NiH$_x$ with $2.0 \leq y \leq 2.4$. The band gap assignment plots are visible in Fig. 8, and the obtained band gaps are summarized in Table 2. The band gap increases for increasing Mg content, in agreement with earlier findings for $a$-Mg$_y$NiH$_x$ films [5]. The band gap tuning is much smaller for crystalline fcc-Mg$_y$NiH$_x$ films. In this work we were able to tune the band gap from 1.54 eV to 1.76 eV for amorphous films and from 2.14 eV to 2.19 eV for crystalline films by variation of the Mg to Ni ratio. The physical origin of the band gap tuning is not well understood [7]. The effect could be convenient for applications of semiconducting Mg-Ni-H in optoelectronics, e. g. solar cells. The modeled gradient in the dielectric functions has negligible influence on the value of the band gap.

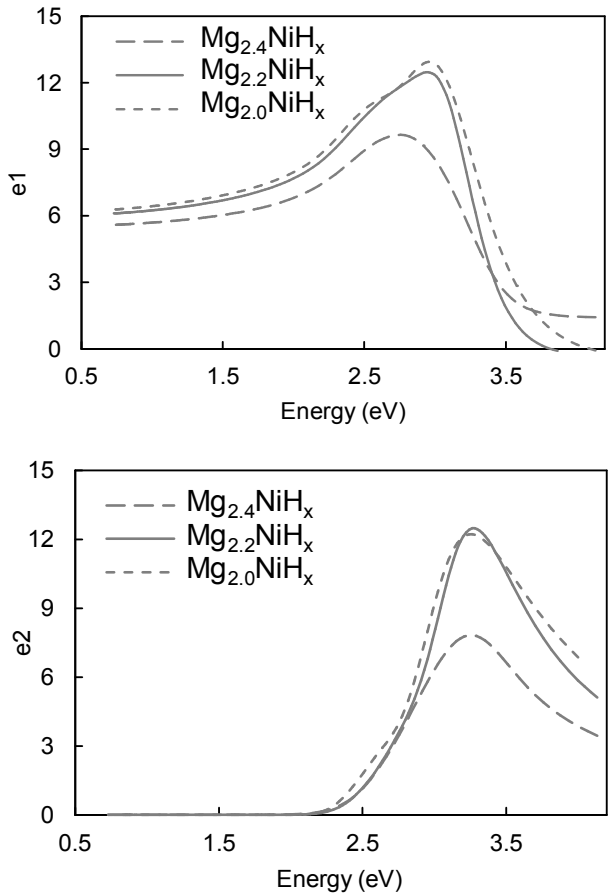

**Fig. 7**. The modeled dielectric functions determined from ellipsometry for annealed c-$Mg_y NiH_x$ samples.

**Table 2.** Band gaps calculated by the Tauc method

| Sample | Bandgap (eV) |
|---|---|
| a-$Mg_{1.9}NiH_x$ | 1.54 |
| a-$Mg_{2.1}NiH_x$ | 1.66 |
| a-$Mg_{2.3}NiH_x$ | 1.76 |
| c-$Mg_{2.0}NiH_x$ | 2.14 |
| c-$Mg_{2.2}NiH_x$ | 2.17 |
| c-$Mg_{2.4}NiH_x$ | 2.19 |

Temperature treatment induces permanent changes in the electrical and optical properties, even when operating below the crystallization temperature. The room-temperature resistivity and the optical transparency increases after annealing at lower temperatures. By using ellipsometry the band gap in an $Mg_{\sim 2}NiH_x$ sample that was treated at 453 K for 1 h was determined to be approximately 1.9 eV. The sample did not show any structural change observable by XRD, and was thus still amorphous after the heat treatment. This observation suggests that parameters like crystallite size and structural defects could have significant impact on the band gap and electronic structure of the films, and this makes it difficult to conclude regarding the actual band gap of amorphous and monoclinic $Mg_2NiH_4$.

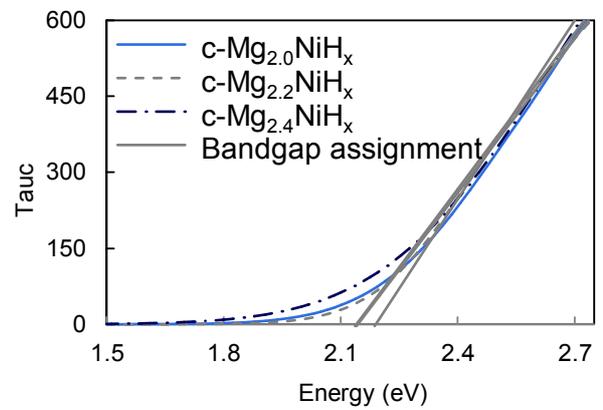

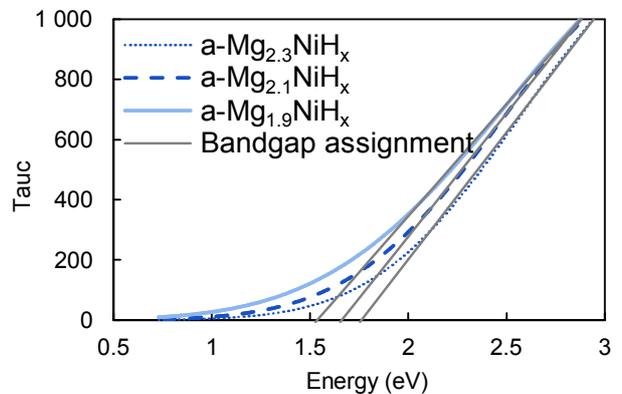

**Fig. 8.** Estimation of the optical band gap for $Mg_y NiH_x$ films based on the Tauc method.

## 5 Conclusions

The optical properties of thin films of amorphous and fcc crystalline $Mg_y NiH_x$ films with $1.9 \leq y \leq 2.4$ have been evaluated. The band gap of amorphous and crystalline films have been estimated using Tauc analysis, and we conclude that $E_g$ = 1.6 eV for as-deposited $a$-Mg$_2$NiH$_4$ and $E_g$ = 2.1 for fcc $c$-Mg$_2$NiH$_4$. Heat treatment of the as deposited $a$-Mg$_2$NiH$_4$ films at low temperatures resulted in an increased band gap, but films of crystalline monoclinic LT-Mg$_2$NiH$_4$ could not be formed. However, TEM analysis has indicated that the amorphous films consist of nano-crystalline monoclinic domains. Based on the experience with heat treatment below the HT crystallization temperature, we believe that crystalline LT-Mg$_2$NiH$_4$ with more long-range ordering has a higher band gap than 1.6 eV.

## Acknowledgements

This work has been funded by the Research Council of Norway through the project "Thin and highly efficient silicon-based solar cells incorporating nanostructures," NFR Project No. 181884/S10. We thank Göran Possnert and Max Wolff, Uppsala University, Sweden, for help with the RBS and NRA measurements.

## References

[1]  J.N. Huiberts, R. Griessen, J.H. Rector, R.J. Wijngaarden, J.P. Dekker, D.D. Groot, N.J. Koeman, Nature 380 (1996) 231-234.
[2]  T. Richardson, J.L. Slack, R. Armitage, R. Kostecki, B. Farangis, M.D. Rubin, Appl. Phys. Lett. 78 (2001) 3047.
[3]  S.Z. Karazhanov, A.G. Ulyashin, EPL 82 (2008) 48004.
[4]  M. Lelis, D. Milcius, E. Wirth, U. Hålenius, L. Eriksson, K. Jansson, K. Kadir, J. Ruan, T. Sato, T. Yokosawa, D. Noreus, J. Alloys Compd. 496 (2010) 81-86.
[5]  R. Gremaud, J.L.M.V. Mechelen, H. Schreuders, M. Slaman, B. Dam, R. Griessen, Int. J. Hydrogen Energy 34 (2009) 8951-8957.
[6]  J. Ell, a Georg, M. Arntzen, a Gombert, W. Graf, V. Wittwer, Sol. Energy Mater. Sol. Cells 91 (2007) 503-517.
[7]  W. Lohstroh, R.J. Westerwaal, J.L.M. van Mechelen, H. Schreuders, B. Dam, R. Griessen, J. Alloys Compd. 430 (2007) 13-18.
[8]  D. Lupu, R. Sârbu, A. Biriş, Int. J. Hydrogen Energy 12 (1987) 425-426.
[9]  W.R. Myers, L.-W. Wang, T. Richardson, M.D. Rubin, J. Appl. Phys. 91 (2002) 4879.
[10] J. Isidorsson, I.A.M.E. Giebels, M.D. Vece, R. Griessen, in:, Proc. SPIE Vol. 4458, 2001, pp. 128-137.
[11] W. Lohstroh, R.J. Westerwaal, J.L.M.V. Mechelen, C. Chacon, E. Johansson, B. Dam, R. Griessen, Phys. Rev. B 70 (2004) 165411.
[12] R.J. Westerwaal, M. Slaman, C.P. Broedersz, D. Borsa, B. Dam, R. Griessen, A. Borgschulte, W. Lohstroh, B. Kooi, G.T. Brink, K.G. Techerisch, H.P. Fleischhauer, J. Appl. Phys. 100 (2006) 063518.
[13] J. Tauc, Amorphous and Liquid Semiconductors, Plenum Press, London, 1974, p. 159-214 .
[14] W.A. Lanford, H.P. Trautvetter, J.F. Ziegler, J. Keller, Appl. Phys. Lett. 28 (1976) 566-568.
[15] M. Mayer, SIMNRA User's Guide (Report IPP 9/113), Garching, Germany, 1997.
[16] G.E. Jellison, F.A. Modine, Appl. Phys. Lett. 69 (1996) 371-373.
[17] D. Noreus, L. Kihlborg, J. Less-Common Met. 123 (1986) 233-239.
[18] K. Yvon, J. Schefer, F. Stucki, Inorg. Chem. 20 (1981) 2776-2778.
[19] G.N. García, J.P. Abriata, J.O. Sofo, Phys. Rev. B 59 (1999) 11746-11754.
[20] T. Mongstad, C.C. You, A. Thøgersen, J.P. Maehlen, C. Platzer-Björkman, B.C. Hauback, S.Z. Karazhanov, Submitted to J. Alloys Compd. (2012).
[21] E. Palik, Handbook of Optical Constants of Solids, Elsevier, 1998, p. 950-951.